\documentclass[fleqn,12pt]{wlscirep}
\usepackage{times}
\usepackage[OT1]{fontenc} 
\usepackage{color}
\usepackage{ragged2e}
\usepackage[version=4]{mhchem}
\usepackage[labelfont=bf,justification=justified]{caption}

\title{\rmfamily Chirality Induced Spin Selectivity in Chiral Crystals}

\author[1,2,$\dagger$]{\rmfamily Qun Yang}
\author[2,$\dagger$]{\rmfamily Yongkang Li}
\author[1]{\rmfamily Claudia Felser}
\author[2,*]{\rmfamily Binghai Yan}
\affil[1]{\rmfamily Max Planck Institute for Chemical Physics of Solids, 01187 Dresden, Germany.}
\affil[2]{\rmfamily Department of Condensed Matter Physics, Weizmann Institute of Science, Rehovot 7610001, Israel}
\affil[$\dagger$]{\rmfamily These authors contributed equally}

\affil[*]{\rmfamily binghai.yan@weizmann.ac.il}

\cleardoublepage

\begin{abstract}
\rmfamily
Chirality is a fundamental property of great importance in physics, chemistry, and biology, and is recently found to generate unexpected spin polarization for electrons passing through organic molecules, known as chirality-induced spin selectivity (CISS). 
CISS shows promising application potential in spintronic devices, spin-controlled chemistry, and enantiomer separation. It focuses on organic molecules that exhibit poor electronic conductivity and inherent complexities, such as the debated role of SOC at the molecule-metal interface.
In this work, we go beyond organic molecules and study chiral solids with excellent electrical conductivity, intrinsic SOC, and topological electronic structures. We demonstrate that electrons exhibit both spin and orbital polarization as they pass through chiral crystals. Both polarization increases with material thickness before saturating to the bulk values. 
The spin polarization is proportional to intrinsic SOC while the orbital polarization is insensitive to SOC. The large spin polarization comes with strong electrical magnetochiral anisotropy in the magneto-transport of these chiral crystals (e.g., RhSi). Our work reveals the interplay of chirality, electron spin, and orbital in chiral crystals, paving the way for developing chiral solids for chirality-induced phenomena.
\end{abstract}
\begin{document}

\flushbottom
\maketitle
\thispagestyle{empty}

\section*{INTRODUCTION}
Chirality, which refers to a unique spatial asymmetry with non-superimposable mirror images, is a fundamental concept with far-reaching implications in the realms of chemistry, biology, and physics \cite{siegel1998homochiral}. Recently, structural chirality was found to induce electronic chirality, manifesting as orbital-momentum locking encoded in the wave function of chiral molecules or solids \cite{liu2021chirality, yang2023monopole,yan2023structural}. Furthermore, the combination of structural chirality with topological band structures leads to chiral multi-fold fermion crossings characterized by parallel spin-momentum locking (\textbf{Fig. 1a}) \cite{sessi2020handedness, chang2018topological, YaoNatCommunObservation, krieger2022parallel}. These findings reveal a deep connection between chirality in the real and reciprocal space of quantum materials, providing an ideal playground for the study of exotic physical phenomena. One of the intriguing phenomena known as chirality-induced spin selectivity (CISS), in which electrons with particular spin can pass through chiral materials (\textbf{Fig. 1b}), generates remarkable spin polarization without the need for magnetic fields or magnetism \cite{naaman2019chiral, naaman2020chiral}. This particular feature of the CISS effect is not only of fundamental importance for understanding the underlying mechanism of electron dynamics in chiral objects but also offers practical promise for various applications, including the field of spintronic devices\cite{yang2021chiral, naaman2019chiral, naaman2015spintronics}, spin control chemistry \cite{zhang2018enhanced, mtangi2017control, bian2022hybrid, naaman2020chiral} and enantiomer separation\cite{banerjee2018separation, stolz2022asymmetric}.

To date, most CISS studies have been carried out on chiral molecules\cite{gohler2011spin,nino2014enantiospecific,mondal2021spin}, self-assembled chiral molecular films\cite{ray1999asymmetric,ziv2019afm}, metal-organic frameworks\cite{huizi2020ideal}, chiral organic-inorganic hybrid perovskites\cite{lu2019spin,
kim2021chiral} and chiral molecular intercalation superlattices\cite{qian2022chiral,al2022atomic}. 
In these systems, organic chiral molecules bring chirality but also complexity, such as the debating role of spin-orbit coupling (SOC) at the molecule-metal interface \cite{liu2021chirality, Adhikari2023,Gersten2013}. 
In contrast, inorganic crystals with homochiral crystal structures and rich electronic structures open up a range of new possibilities for the CISS effect. 
Recently, it was observed that chiral crystals of Te\cite{calavalle2022gate,furukawa2017observation}, \ce{CrNb3S6}\cite{inui2020chirality} and \ce{TMSi2} (TM = Nb, Ta)\cite{shiota2021chirality} exhibit a spin-polarized state when subjected to charge current injection. Theoretically, the physical origin of the observed phenomenon remains elusive. Existing theories have endeavored to provide an explanation by linking it to the Edelstein effect that emerges from the radial spin texture and spin accumulation~\cite{calavalle2022gate,Slawinska2022,Slawinska2023}. 
However, the Edelstein effect refers to uniform magnetization inside the crystal induced by a non-spin-polarized current\cite{yang2023monopole, zhong2016gyrotropic, xiao2010berry} and is not fully equivalent to the spin-polarized current generated by electrons transporting through the chiral material. 

In this work, we perform quantum transport calculations on chiral crystals and evaluate the spin polarization of current induced by the chirality. We find that electrons transmitted through these chiral crystals exhibit both spin and orbital polarization. Both polarizations align along the transport direction and increase with the propagating distance before saturating when the system approaches bulk properties, consistent with the length-enhanced spin polarization observed in chiral molecules. Comparing various compounds with the same structural and electronic properties, we reveal that induced spin polarization is proportional to the SOC magnitude while orbital polarization is insensitive to SOC. Furthermore, we find a spin polarization ratio of up to 40\% in Te, originating from the large spin splitting at the valence band edge. The pronounced spin polarization leads to a substantial electric magneto-chiral anisotropy  (EMCA) when probed by magnetic electrodes. We demonstrate that RhSi exhibits a magneto-current ratio (0.22\%) upon flipping the electrode magnetization, which is larger than the relative change of magneto-resistance observed in earlier experiments. The distinction between the spin polarization and magneto-current ratio is emphasized in our analysis. Our results pave the way to understanding the CISS effect in inorganic crystals and explore emerging chirality-driven phenomena. 

\begin{figure}[htbp]
    \centering
    \includegraphics[width=0.8\textwidth]{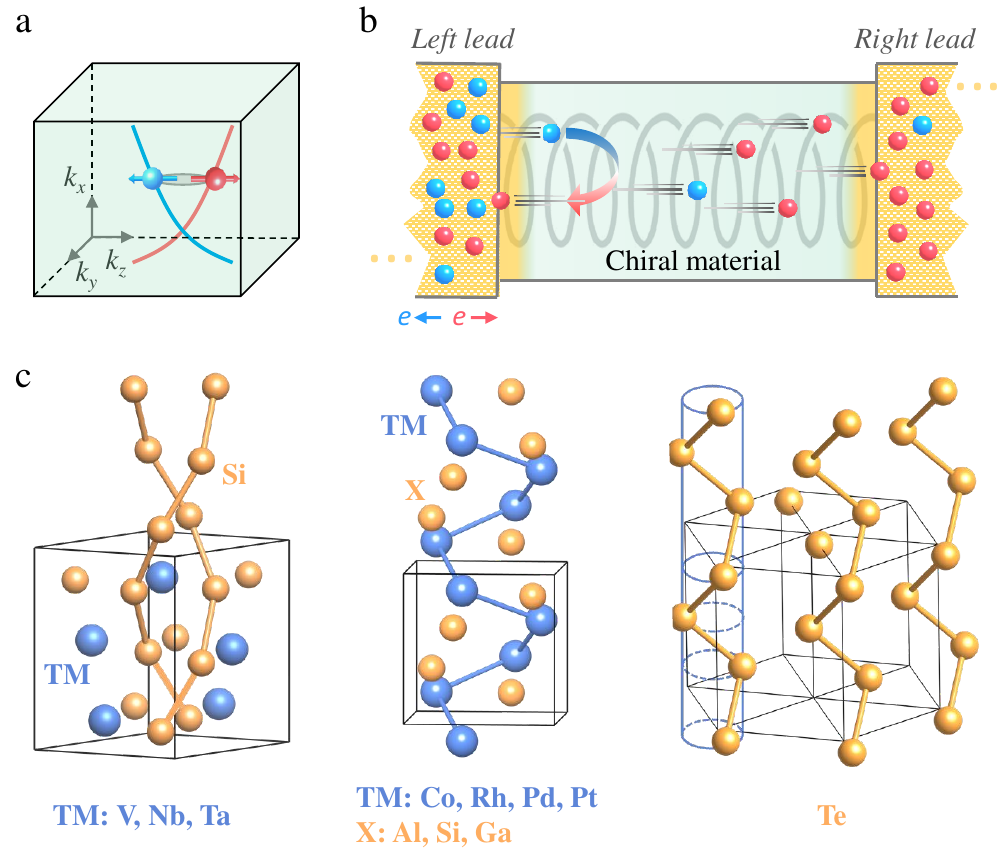}
    \caption{\textbf{Schematic illustration of the two-terminal device setup for the chirality-induced spin selectivity 
    (CISS) effect in chiral crystal and crystal structures of representative chiral materials.} 
\textbf{a.} Illustration of the electronic chirality of spin-momentum locking on the band structure of chiral materials. \textbf{b}. Geometry of the transport model. The chiral crystal is placed in the central region, which is periodically 
    repeated in the transverse direction but finite in the transport direction. The chiral crystal seamlessly 
    connects with two leads. The electrons with opposite spins (red and blue spheres) moving through the chiral central region 
    become spin-polarized. The chiral crystal acts as the spin polarizer so that the transmitted and reflected electrons exhibit 
    the same type of spin polarization. 
\textbf{c}. Chiral crystal structures of the transition metal disilicide \ce{TMSi2} (TM = V, Nb, Ta), FeSi-type B20 materials TMX (TM = Co, Rh, Pd, Pt; X = Al, Si, Ga), and trigonal tellurium (Te). Their chirality can be 
    distinguished by the helical chains formed by Si, TM, and Te atoms, respectively}
    \label{Figure1}
\end{figure}

\section*{RESULTS AND DISCUSSION}
Among the 230 space groups (SGs), 65 are designated as Sohncke SGs for chiral structures. Within these, 22 are enantiomeric, forming 11 enantiomorphous pairs, such as P6$_1$ and P6$_5$. Additionally, there are 43 non-enantiomeric SGs, each containing two enantiomers that share identical spatial symmetry but inverted atomic positions, such as P6$_3$ and P2$_1$3. In our research, we consider three catalogs of representative chiral crystals, including 1) Intermetallic transition metal disilicides, specifically \ce{TMSi2} (TM=V, Nb, and Ta). These crystals adopt C40 structures and belong to the enantiomeric SG P6$_2$22 (No. 180) or SG P6$_4$22 (No. 181). 2) Topological multifold semimetals, in particular, TMX compounds (TM = Co, Rh, Pd, Pt, and X = Al, Si, Ga). These materials crystallize in B20 (FeSi type) structures and belong to the non-enantiomeric SG P2$_1$3 (No. 198). The topological trivial insulator RhBiS is also included. 3) Semiconductor $\gamma$-Se and Te, characterized by a trigonal crystal system and belonging to the enantiomeric SG P3$_1$21 (No. 152) or P3$_2$21 (No. 154). \textbf{Fig. 1c} shows the crystal structures of these chiral materials in one of their enantiomeric forms where the chirality can be visualized by the helical winding patterns of the atoms. 

To investigate the dynamic CISS process in these chiral crystals, we constructed a two-terminal CISS device for quantum transport calculations, as depicted in \textbf{Fig. 2a}. The chiral crystal with finite thickness along the [001] transport direction is positioned on the central region, which is connected with two semi-infinite leads on its left and 
right sides. In both leads, we employed the wide-band limit (WBL) approximation \cite{ryndyk2016theory,covito2018transient}, assuming a constant density of states (DOS) that 
disregards the detailed structure of the DOS in the leads. This approximation is commonly used to describe transport in 
nanoscale devices, which substantially simplifies computations. Within the WBL framework, the spin and orbital polarization can 
be accurately defined within the leads. The inherent SOC within the chiral crystals facilitates preferential spin/orbital 
polarization of the transmitted electrons, depending on the current direction and handedness of the chiral crystal. 
Employing this model, we calculated the conductance (G) using the Landauer-B{\"u}ttiker formalism\cite{buttiker1986four}, which involves the transmission function from left (L) to the right (R) lead,  
\begin{equation} \label{Landauer-Buttiker formalism}
     \rm G_{RL} = \frac{e^2}{h}\sum\nolimits_{n\in R,m\in L}|S_{nm}|^2
\end{equation}
\noindent where S$_{\rm nm}$ is the scattering matrix element from the $m$th eigenstate in L lead to the $n$th eigenstate in R lead. With the spin and orbital conserved leads, the spin- and orbital-polarized conductance can be obtained as
\begin{equation} \label{spin conductance}
     \rm G_{RL}^{S_z} = \frac{e^2}{h}\sum\nolimits_{n\in R,m\in L}S_{z,n}|S_{nm}|^2
\end{equation}
\vspace{-8ex}
\begin{equation} \label{orbital conductance}
     \rm G_{RL}^{L_z} = \frac{e^2}{h}\sum\nolimits_{n\in R,m\in L}L_{z,n}|S_{nm}|^2
\end{equation}

\begin{figure}[htbp]
    \centering
    \includegraphics[width=0.9\textwidth]{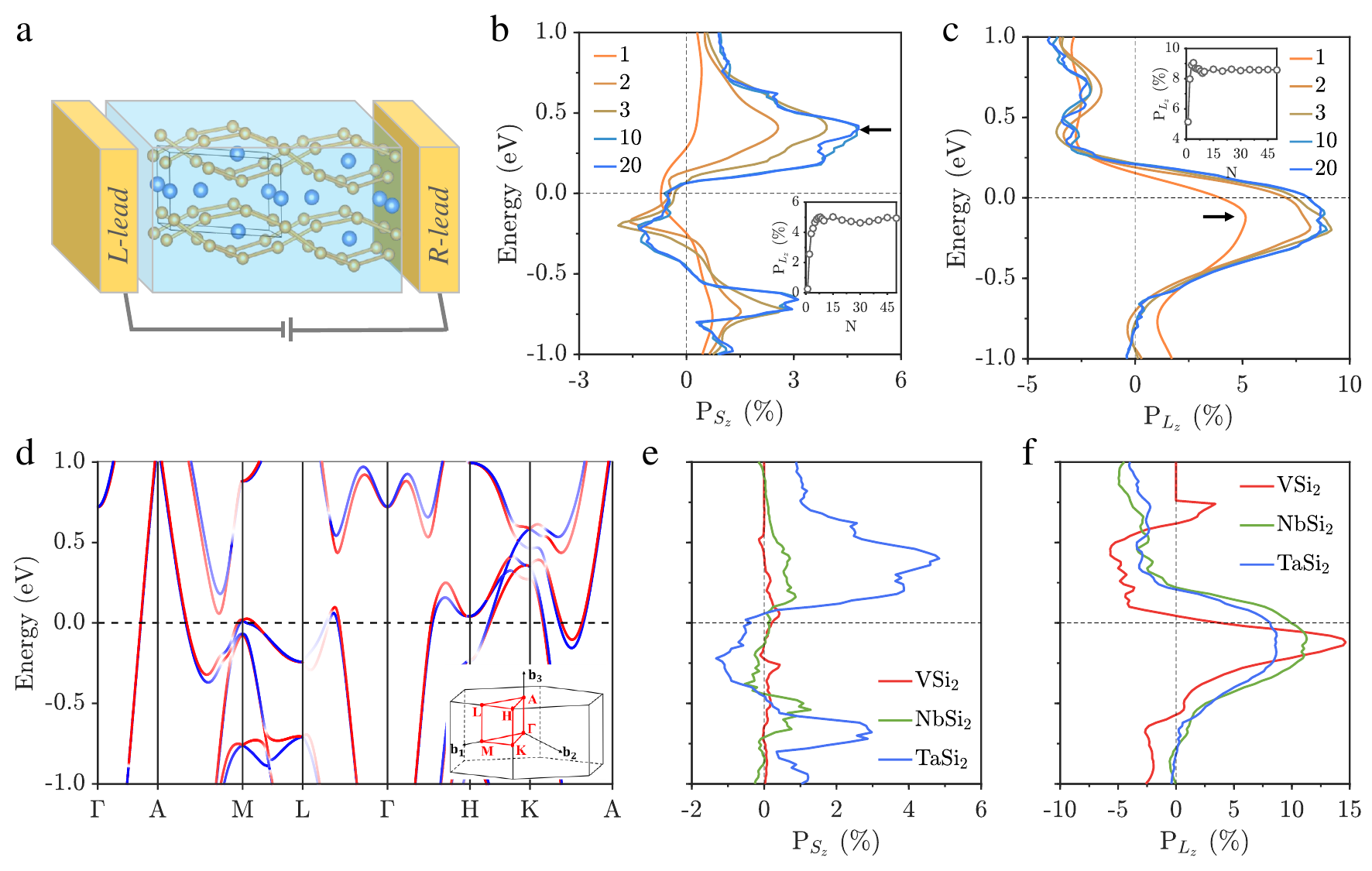}
    \caption{\textbf{Spin polarization and orbital polarization in the conductance of transition metal 
    disilicide \ce{TMSi2} compounds.} 
    \textbf{a}. The transport model consists of two leads and the chiral material \ce{TaSi2} with thickness NL along the [001] 
    transport direction in the central region. N and L denote the number of unit cells and the lattice constant along the crystal c-axis. N=2 is shown in schematic.
    \textbf{b}. Energy-dependent spin polarization ratio (P$\rm _{S_z}$) and \textbf{c}. orbital polarization ratio (P$\rm _{L_z}$) of \ce{TaSi2} for N=1, 2, 3, 10 and 20. Peaks of P$\rm _{S_z}$ and P$\rm _{L_z}$ (see inset in \textbf{b} and \textbf{c}, respectively) increase rapidly with increasing N and approach saturation when N reaches 10. For all subsequent calculations, N is fixed at 20 to fundamentally explore the intrinsic bulk band structure effect on spin and orbital polarization. 
    \textbf{d}. Spin angular momentum (SAM) resolved band structures of \ce{TaSi2} in the Brillouin zone. The color red (blue) 
    represents spin-up (down) states. Energy-dependent 
    \textbf{e}. P$\rm _{S_z}$ and \textbf{f}. P$\rm _{L_z}$ for \ce{VSi2}, \ce{NbSi2}, and \ce{TaSi2}, respectively. The Fermi energy lies at the charge neutral point.}
    \label{Figure2}
\end{figure}

\noindent G$\rm _{RL}^{S_z(L_z)}$ is the conductance from L lead to the S$\rm _z(L_z)$ channel of R lead. S$\rm _{z,n}(L_{z,n})$ is the z-component of spin (orbital) angular momentum carried by the transmitted electron in mode n of the R lead. 
The corresponding spin- and orbital-polarization ratio can be calculated by P$\rm _{S_z}=\frac{G_{S_z}}{G}$$\times$ 100\% 
and P$\rm _{L_z}=\frac{G_{L_z}}{G}$$\times$ 100\%, respectively. In the following discussion, we use P$\rm _{S_z(L_z)}$, 
G$\rm _{S_z(L_z)}$, and G for P$\rm _{RL}^{S_z(L_z)}$, G$\rm _{RL}^{S_z(L_z)}$, and G$\rm _{RL}$, respectively, if it is not noted specifically. 

Experimental observations of the CISS effect on chiral molecules reveal a notable correlation between the induced spin polarization and the molecular length\cite{kettner2015spin,gohler2011spin}. To explore this phenomenon in chiral crystal, we initially examined the dependence of the P$\rm _{S_z}$ and P$\rm _{L_z}$ on the thickness of chiral material \ce{TaSi2} in the central scattering region, quantified by the number of unit cells (N) in the transport direction (the [001] direction). As shown in \textbf{Fig. 2b-c}, the results confirm the coexistence of P$\rm _{S_z}$ and P$\rm _{L_z}$, attributing to the presence of the intrinsic SOC and inversion symmetry breaking within the chiral material. Especially, the P$\rm _{S_z}$ exhibits an initial increase followed by a saturation when N reaches ten, indicating a thickness-enhanced spin polarization in chiral crystal with saturation occurring when the system reaches a bulk characteristic. While P$\rm _{L_z}$ exhibits an initial increase until N=4, followed by a slight decrease, and then reaches saturation at around N=10. Moreover, we point out that the chiral crystal acts as the spin polarizer(also orbital polarizer) \cite{wolf2022unusual} , resulting in the transmitted and reflected electrons exhibiting the same type of spin polarization (orbital polarization), as confirmed by our transmitted/reflected spin (orbital) conductance calculations. Furthermore, P$\rm _{S_z}$ and P$\rm _{L_z}$ reverse its sign for opposite enantiomers A and B: P$\rm _{S_z}^A$=-P$\rm _{S_z}^B$; P$\rm _{L_z}^A$=-P$\rm _{L_z}^B$.

To understand the origin of the chirality-induced spin and orbital polarization in chiral materials, the spin angular momentum (SAM)- and orbital angular momentum (OAM)-resolved band structures of \ce{TaSi2} are further analyzed. \ce{TaSi2} in SG P6$_2$22 possesses threefold screw rotation $\{C_{3z} | 0, 0, 2/3\}$ and twofold rotations $\{C_{2,110} | 0, 0, 2/3\}$ and $\{C_{2z} | 0, 0, 0\}$ as the symmetry generators. These symmetries impose constraints on the SAM and OAM in the momentum space. We found that the z-component of SAM and OAM must vanish along all high symmetry lines in the k$\rm _z=0$ and k$\rm _z=\pi$ planes. As an illustration, the high-symmetry line $\Gamma$-K (k$_x$=k$_y$=k, k$_z$=0) remains invariant under the C$_{2,110}$ axis, resulting in L$\rm  _z (k,k,0)=0$. Therefore, the SAM- and OAM-resolved band structures are computed for \ce{TaSi2} along the selected k-paths where S$\rm _z (k)$ and L$\rm _z (k)$ are non-zero. As shown in \textbf{Fig. 2d} and \textbf{Fig. S1}, the chiral crystalline symmetries and strong SOC in \ce{TaSi2} result in a large spin-splitting of the bands. We found that the spin-split bands carry the opposite SAM but approximately the same OAM. This induces the spin and orbital polarization as electrons move through the chiral crystal and generally a larger magnitude of P$\rm _{L_z}$ than P$\rm _{S_z}$ in the presence of current flow. At the charge neutral point, the calculated P$\rm _{S_z}$ and P$\rm _{L_z}$ in \ce{TaSi2} is -0.57\% and 7.95\%, respectively. 

\textbf{Fig. 2e} and \textbf{Fig. 2f} shows the energy-dependent P$\rm _{S_z}$ and P$\rm _{L_z}$ for \ce{TaSi2}-family of materials, including \ce{TaSi2}, \ce{NbSi2}, and \ce{VSi2}. We found that the P$\rm _{L_z}$ is insensitive to the SOC strength of different materials,  reaching the maximum value of 14.60\% at -0.10 eV in \ce{VSi2}. In contrast, P$\rm _{S_z}$ shows a clear correlation with the SOC strength in materials. The peaks of P$\rm _{S_z}$ appear at positions with pronounced spin splitting in the band structure. Especially, within a wide energy range, the P$\rm _{S_z}$ increases as SOC strength increases in V-3d, Nb-4d and Ta-5d\cite{onuki2014chiral}. For instance, the calculated P$\rm _{S_z}$ at peak 0.21 eV are 0.08\%, 0.62\%, and 3.83\% in \ce{VSi2}, \ce{NbSi2}, and \ce{TaSi2}, respectively. The maximum P$\rm _{S_z}$ value of 4.84\% is observed at 0.41 eV for \ce{TaSi2}. 
It’s noted that the CISS effect has predominantly been explored in chiral organic molecules with P$\rm _{S_z}$ values exceeding 60\%\cite{al2022atomic, qian2022chiral, naaman2020chiral}. By contrast, the magnitude of P$\rm _{S_z}$ in chiral crystals is small because the giant total conductance in these metals. Despited of the smallness of P$\rm _{S_z}$, the excellent electrical conductivity with high carrier density in these chiral metals can generate a large amount of spin-polarized carriers. Thus, we expect that the CISS in the chiral crystal is promising for application in spintronic devices, spin-selective chemical reactions, and enantiomer recognition, which remains poorly explored in chiral inorganic crystals\cite{inui2020chirality, shiota2021chirality, shishido2021detection}. 


\begin{figure}[htbp]
    \centering
    \includegraphics[width=0.9\textwidth]{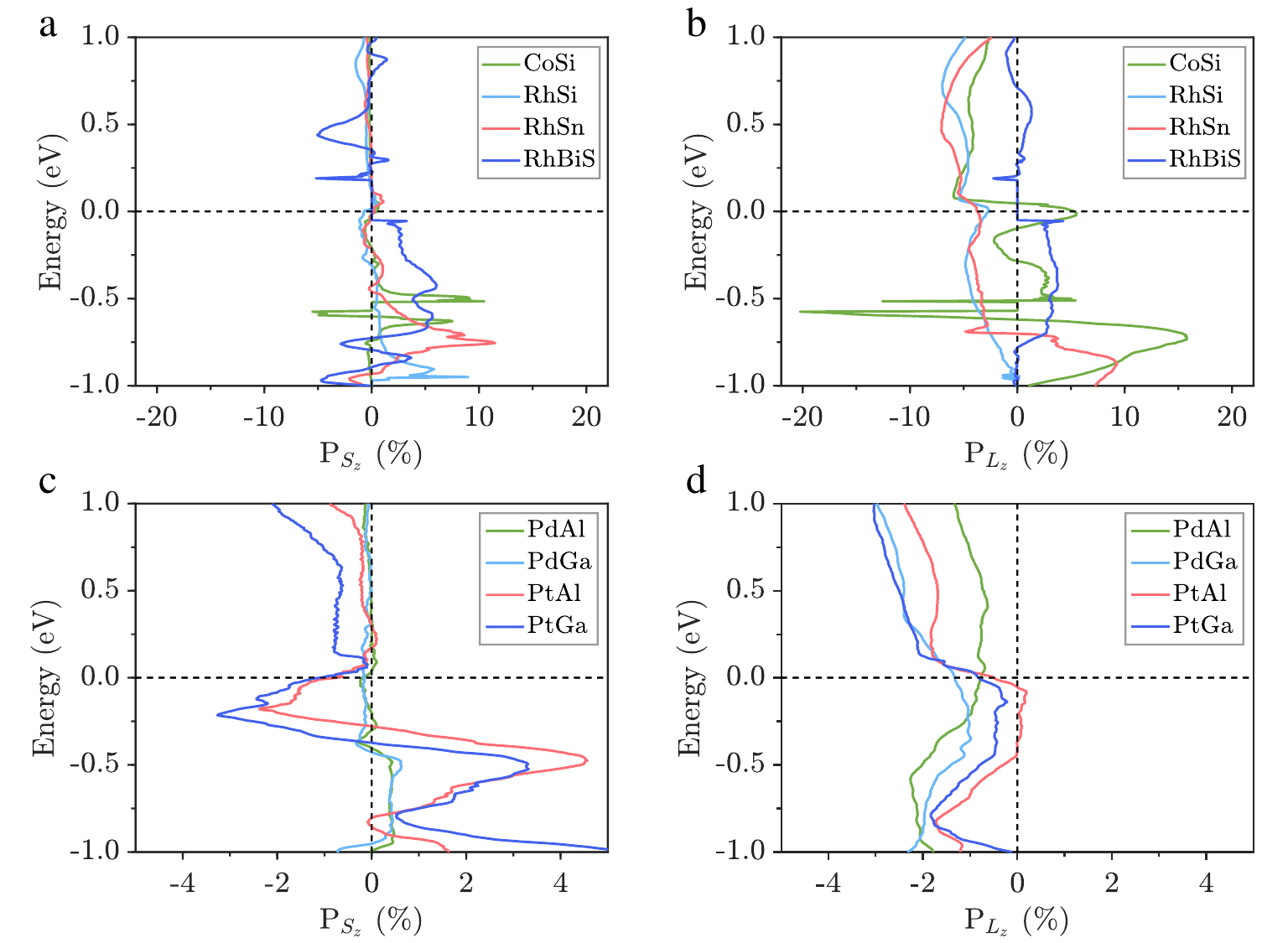}
    \caption{\textbf{Spin and orbital polarization in the conductance of some representative chiral materials 
    with the SG P2$_1$3.} 
    The graph displays the values of P$\rm _{S_z}$ and P$\rm _{L_z}$ for topological chiral semimetals within the 
    FeSi-type B20 class, including members from the RhSi family (CoSi, RhSi, and RhSn), as well as the PdGa family of 
    materials (PdAl, PdGa, PtAl, and PtGa). Additionally, the figure includes data for the semiconducting chiral 
    material, RhBiS.}
    \label{Figure3}
\end{figure}

Next, we studied the CISS effect within the family of chiral materials belonging to SG P2$_1$3. \textbf{Fig. 3} illustrates the calculated P$\rm _{S_z}$ and P$\rm _{L_z}$ of various representative topological chiral semimetals within this SG, including the RhSi family (CoSi, RhSi, and RhSn) and PdGa family (PdAl, PdGa, PtAl, and PtGa), based on their elemental similarities. These materials are characterized by multiple band crossings with large Chern numbers in the bulk state and unique Fermi arcs at surfaces\cite{YangAdvMaterTopological, 
NielsScienceObservation,  YaoNatCommunObservation, RaoNatureObservation}. The z-component of their SAM- and OAM-resolved band structures are displayed in \textbf{Fig. S2} and \textbf{Fig. S3}. We found that the P$\rm _{S_z}$ values of the RhSi family of materials do not exhibit a clear trend. The sharp peaks observed at -0.52 eV and -0.58 eV in CoSi, -0.91 eV in RhSi, and -0.71 eV in RhSn primarily arise from the low total conductance at the band edge. Notably, RhBiS is identified as a topological trivial semiconductor\cite{zhang2019strong} with an indirect band gap of approximately 0.25 eV by our GGA-PBE calculation. The top of its occupied bands is mainly dominated by the Bi-6$p$ orbital, leading to significant spin splitting. We found that the spin polarization disappears within the band gap, but it exhibits a relatively large value of about 5\% across a wide energy range spanning from -0.6 eV to -0.1 eV, surpassing that of CoSi, RhSi, and RhSn. For the PdGa family of materials, the PtAl and PtGa exhibit evident enhanced P$\rm _{S_z}$ values compared to PdAl and PdGa. At the energy of -0.48 eV, where the peak appears, the calculated P$\rm _{S_z}$ values are 0.40\%, 0.59\%, and 4.53\%, 3.18\% for PdAl, PdGa, PtAl and PtGa, respectively. This is probably because Pt atoms have larger atomic numbers and hence stronger SOC effects than Pd atoms. Recent studies reveal that the topological chiral semimetals such as PdGa show promising applications in enantioselective catalytic reactions of chiral molecules\cite{prinz2015highly, stolz2022asymmetric, li2023observation}, which is a signature of the existence of the exotic interplay between the structural chirality of the material and its electron spin or orbital transport properties\cite{yang2023monopole}. Our results reveal that the presence of the specific spin and orbital polarization carried by the transmitted electrons through PdGa will facilitate the theoretical understanding of the enantioselective chemistry in the material\cite{wang2023direct}.

Generally, semiconductors or insulators, with their significantly lower carrier density, will show a larger magnitude of spin polarization compared to metals. The $\gamma-Se$ is a semiconductor with a large indirect band gap of 1.058 eV by our GGA-PBE calculations, where the valence band maximum and conduction band minimum are located at the L and H points of the Brillouin Zone (BZ), respectively, as shown in \textbf{Fig. 4a}. We found that P$\rm _{S_z}$ is smaller than P$\rm _{L_z}$ in $\gamma-Se$ with the amplitude of a few percent near the Fermi level. Particularly, Te is also a semiconductor, but with a direct band gap of 0.142 eV formed at the H points. The strong spin-orbit interaction in Te causes a substantial separation of 0.126 eV between the two upper branches of the valence band at the H point, significantly impacting its electrical transport properties. We found that the P$\rm _{L_z}$ is greater than P$\rm _{S_z}$ in the conductance band region. However, the trend is reversed in the valence band region with P$\rm _{S_z}$ typically dominating over P$\rm _{L_z}$ (\textbf{Fig. 4d}). Strikingly, a sharp peak with a giant P$\rm _{S_z}$ of approximately 40\% is observed around the valence band edge. This high P$\rm _{S_z}$ is attributed to the single band occupancy, as the closest band in energy with an opposite SAM is largely separated with the valence band edge (\textbf{Fig. 4c}). This situation is ideal for maximizing the P$\rm _{S_z}$ and, consequently, the CISS effect. We point out that the giant P$\rm _{S_z}$ observed in Te is comparable to the value reported in chiral organic molecules. The large polarization originates in the large spin splitting and small charge conductance at the band edge of chiral semiconductors, compared to the above chiral semimetals/metals. 

\begin{figure}[htbp]
    \centering
    \includegraphics[width=0.9\textwidth]{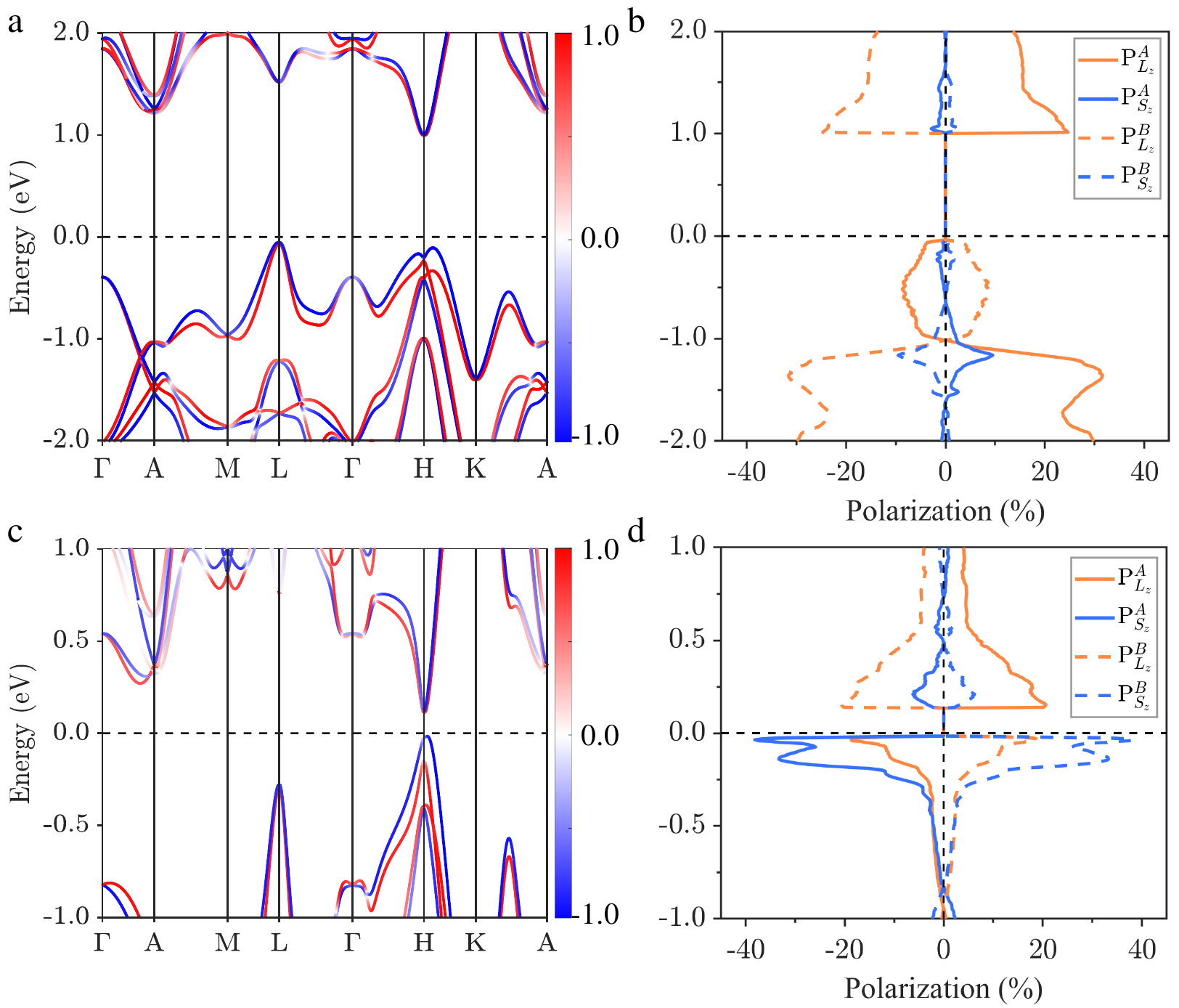}
    \caption{\textbf{Spin polarization and orbital polarization in the conductance of $\gamma$-Se and Te.} SAM-resolved band structures of the enantiomer A of \textbf{a.} $\gamma$-Se and \textbf{c.} Te. The color red (blue) represents spin up (down) states.  $\rm P_{S_z}$ and $\rm P_{L_z}$ in enantiomer A ($\rm P_{S_z}^A$, $\rm P_{L_z}^A$) and B ($\rm P_{S_z}^B$, $\rm P_{L_z}^B$) of \textbf{b.} $\gamma$-Se and \textbf{d.} Te.
    }
    \label{Figure4}
\end{figure}

In experiments, CISS is commonly probed through two-terminal magnetoresistance (MR) measurements, with a chiral molecule sandwiched between a ferromagnetic substrate and a nonmagnetic lead\cite{qian2022chiral,liu2020linear,lu2019spin}. The induced spin polarization from chiral molecules leads to resistance changes when flipping the substrate magnetization (M). We highlight that in the context of CISS, the induced spin polarization can also lead to electrical magneto-chiral anisotropy (EMCA) transport\cite{rikken2001electrical,liu2021chirality}, which is usually reported in chiral solids. It refers to the resistance that depends on the handedness of the conductor and on the relative orientation of electrical current I and magnetic field B. The EMCA resistance is described by the equation R$^\chi$=R$_0$(1+$\alpha B^2$+$\beta^\chi B\cdot I$), where $\beta^\chi=-\beta^{-\chi}$ and $\chi=\pm$ represents the chirality. The B$^2$ term represents the ordinary magnetoresistance. Notably, the B$\cdot$I term is specific to chiral conductors, indicating the presence of unidirectional magnetoresistance (UMR): $R^\chi(I, B)\neq R^\chi(I, -B)$. EMCA transport respects the Onsager's reciprocal relation, signifying the presence of microscopic reversibility\cite{rikken2001electrical}. By contrast, CISS MR in chiral molecules violates this reciprocity, leading to a fundamentally distinct symmetry in the I-V curves and conductance compared to EMCA\cite{dalum2019theory, yang2019spin, yang2020detecting, naaman2020comment, yang2020reply,yan2023structural}. Our recent study reveals a close relationship between EMCA and CISS MR, with EMCA predominantly manifesting in chiral conductors\cite{xiao2022nonreciprocal}. Experimentally, such EMCA has been demonstrated in many chiral solids, including bismuth helices\cite{rikken2001electrical}, carbon nanotubes\cite{krstic2002magneto,wei2005magnetic}, bulk organic conductors\cite{pop2014electrical}, tellurium semiconductor\cite{rikken2019strong,calavalle2022gate}, metals with intrinsic magnetic order\cite{yokouchi2017electrical,aoki2019anomalous}, and superconductors\cite{qin2017superconductivity}, in which the relative amplitude of UMR was usually several percent or even less, while the CISS MR can be much larger in some molecular devices \cite{al2022atomic,qian2022chiral}.

\begin{figure}[htbp]
    \centering
    \includegraphics[width=0.9\textwidth]{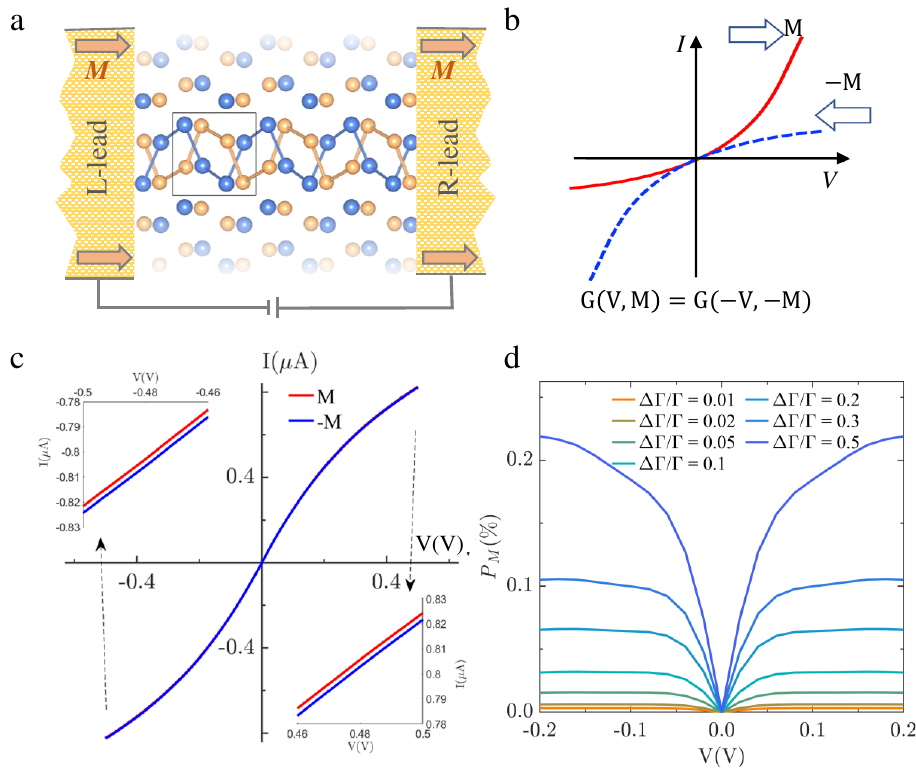}
    \caption{\textbf{Magneto-current in RhSi.} 
    \textbf{a.} Schematic representation of the device setup for electrical magneto-chiral anisotropy (EMCA), with a chiral crystal 
    connecting magnetized leads. The conductance and current are high (low) when the 
    electron spin aligns parallel (or anti-parallel) to the electrode magnetization M. 
    \textbf{b.} Typical current-voltage (I-V) curves for EMCA. The colors represent the electrode magnetization direction.
   \textbf{c.} The I-V curves of EMCA in RhSi. EMCA respects Onsager's reciprocal relation that the magnitude of current in a two-terminal setup remains invariant when the bias voltage (V) and magnetism (M) are reversed, i.e., $\rm I(V, M) = -I(-V, -M)$ as shown in the inset. \textbf{d.} Magneto-current ratio in RhSi. The magneto-current ratio 
    P$\rm _M=\frac{I(V, M)-I(V,-M)}{I(V, M)+I(V,-M)}$$\times$ 100\% increases with increasing magnitudes of $\Delta\Gamma/\Gamma$ which is a representation of magnetization M. 
    } 
    \label{Figure5}
\end{figure}

The spin polarization in CISS is commonly represented by the magnetoresistance ratio of transport measurements. We calculated the magnetoresistance directly and demonstrated that the magnetoresistance is more subtle than the spin polarization. 
We employed the same two-terminal model as in the spin polarization calculation above but introduced a magnetization along the z direction in both leads to mimic the effect of the magnetic field, as depicted in \textbf{Fig. 5a}. Inelastic scattering is taken into account by introducing a dephasing parameter $i\eta$ in the central region. 
This dephasing term breaking the current conservation is necessary to realize the magneto-resistance\cite{liu2021chirality}. The resulting I-V curve of RhSi is shown in \textbf{Fig. 5c}, where spin polarization in the transmission current leads to UMR, i.e. $\rm I(V, M)\neq I(V,-M)$ in the system. But Onsager's reciprocal relation $\rm G(V, M)=G(-V,-M)|_{V\rightarrow 0}$ still holds showing that UMR is mainly caused by EMCA in the chiral conductors. To characterize this UMR effect, the magneto-current ratio is defined as P$\rm _M=\frac{I(V, M)-I(V,-M)}{I(V, M)+I(V,-M)}$$\times$100\%\cite{Liu2023EMCA}. \textbf{Fig. 5d} displays the calculated P$\rm _M$ for RhSi. 
We found that the P$\rm _M$ increases with the magnitude of the magnetization (M), as denoted by the spin-polarized DOS ratio $\Delta\Gamma/\Gamma$ (see Supplementary Information for definition of $\Delta\Gamma/\Gamma$). Especially, at the maximal M ($\Delta\Gamma/\Gamma$=0.5), P$\rm _M$ first increases with bias and then decreases, exhibiting a peak value P$\rm _M=0.22\%$ around $V=\pm200$ meV. Noteworthy, the magnitude of P$\rm _M$ is larger than the observed relative change of magneto-resistance due to EMCA in trigonal tellurium crystals\cite{rikken2019strong}, chiral magnet MnSi\cite{yokouchi2017electrical} and CoNi microhelices\cite{2018Maurenbrecher}, so we expect this effect is observable in RhSi.
We also point out that although related, P$\rm _{S_z}$ is usually different from P$\rm _M$\cite{Liu2023EMCA} as shown in our calculation. Fundamentally, spin polarization is related to chirality and SOC, while magnetoresistance further requires dephasing or dissipation. For example, RhSi has P$\rm _{S_z}\approx 0.7\%$ at Fermi energy but much smaller P$\rm _M<0.25\%$ in the whole bias range. Therefore, detecting CISS via magneto-transport unnecessarily equals the induced spin polarization in chiral crystals.


In summary, we elucidate and quantify CISS and the associated EMCA response in representative chiral crystals by first-principles quantum transport calculations. The combination of intrinsic homochiral crystal structure and SOC in these crystals leads to simultaneous electron spin and orbital polarization. We demonstrate SOC-enhanced spin polarization in chiral crystals with similar structural and electronic properties while orbital polarization remains insensitive to SOC. 
We investigated the chirality-driven magnetoresistance and demonstrated its distinction from the spin polarization. This work provides essential insights into CISS responses in chiral inorganic crystals, paving the way for applications in spintronics, spin control chemistry, and enantiomer recognition. Additionally, we notice a recent theory work reporting giant chirality-induced spin polarization in twisted transition metal dichalcogenides \cite{Menichetti2023}.

\section*{Calculation details}
We performed the density-functional theory (DFT) calculations\cite{kresse1996efficiency, kresse1996efficient} combined with the Full-Potential Local-Orbital (FPLO) package\cite{koepernik1999full} to generate highly symmetric atomic-orbital-like Wannier functions and the corresponding tight-binding (TB) model Hamiltonian of CoSi, RhSi, RhSn, RhBiS, PdAl, PdGa, PtAl, PtGa, and Te. The exchange-correlation potential is described in the generalized gradient approximation (GGA)\cite{perdew1996generalized}. The k-point grid was set to 12$\times$12$\times$12, and the convergence of the total energy convergency was set to 10$^{-6}$ eV. Using the obtained TB Hamiltonian, the quantum transport quantity of the conductance can be calculated based on Landauer-B{\"u}ttiker formulism in Eq. \ref{Landauer-Buttiker formalism}, and the spin and orbital polarized conductance can be calculated by Eq. \ref{spin conductance} and \ref{orbital conductance}. The detailed methods of quantum transport calculations are given in the Supplementary Information.

\section*{Acknowledgement}
Q.Y. thanks Changjiang Yi and Jiewen Xiao for helpful discussions. We acknowledge the financial support by the MINERVA Stiftung, the European Research Council (ERC Consolidator Grant ``NonlinearTopo'', No. 815869), the Israel Science Foundation (ISF, No. 2932/21), the European Research Council (ERC Advanced Grant No.742068), Deutsche Forschungsgemeinschaft (DFG) under SFB 1143 (project ID. 24731007), the Würzburg-Dresden Cluster of Excellence on Complexity and Topology in Quantum Matter—ct.qmat (EXC 2147, project no. 390858490), and DFG (German Research Foundation) for 5249 (QUAST).

\bibliography{reference}

\end{document}